\documentclass[longauth]{aa} % for the long lists of affiliations 
\usepackage{graphicx,natbib}
\usepackage{txfonts}
\begin{document}
   \title{WEBT and XMM-Newton observations of \object{3C 454.3} during the post-outburst phase}

   \subtitle{Detection of the little and big blue bumps\thanks{ The radio-to-optical data 
   presented in this paper are stored in the WEBT archive; for questions regarding their availability,
   please contact the WEBT President Massimo Villata ({\tt villata@oato.inaf.it}).}}

   \author{C.~M.~Raiteri              \inst{ 1}
   \and   M.~Villata                  \inst{ 1}
   \and   V.~M.~Larionov              \inst{ 2,3}
   \and   T.~Pursimo                  \inst{ 4}
   \and   M.~A.~Ibrahimov             \inst{ 5}
   \and   K.~Nilsson                  \inst{ 6}
   \and   M.~F.~Aller                 \inst{ 7}
   \and   O.~M.~Kurtanidze            \inst{ 8}
   \and   L.~Foschini                 \inst{ 9}
   \and   J.~Ohlert                   \inst{10}
   \and   I.~E.~Papadakis             \inst{11,12}
   \and   N.~Sumitomo                 \inst{13}
   \and   A.~Volvach                  \inst{14}
   \and   H~.D.~Aller                 \inst{ 7}
   \and   A.~A.~Arkharov              \inst{ 3}
   \and   U.~Bach                     \inst{15}
   \and   A.~Berdyugin                \inst{ 6}
   \and   M.~B\"ottcher               \inst{16}
   \and   C.~S.~Buemi                 \inst{17}
   \and   P.~Calcidese                \inst{18}
   \and   P.~Charlot                  \inst{19}
   \and   A.~J.~Delgado S\'anchez     \inst{20}
   \and   A.~Di Paola                 \inst{21}
   \and   A.~A.~Djupvik               \inst{ 4}
   \and   M.~Dolci                    \inst{22}
   \and   N.~V.~Efimova               \inst{ 3}
   \and   J.~H.~Fan                   \inst{23}
   \and   E.~Forn\'e                  \inst{24}
   \and   C.~A.~Gomez                 \inst{ 2}
   \and   A.~C.~Gupta                 \inst{25}
   \and   V.A.~Hagen-Thorn            \inst{ 2}
   \and   L.~Hooks                    \inst{16}
   \and   T.~Hovatta                  \inst{26}
   \and   Y.~Ishii                    \inst{13}
   \and   M.~Kamada                   \inst{13}
   \and   N.~Konstantinova            \inst{ 2}
   \and   E.~Kopatskaya               \inst{ 2}
   \and   Yu.~A.~Kovalev              \inst{27}
   \and   Y.~Y.~Kovalev               \inst{15,27}
   \and   A.~L\"ahteenm\"aki          \inst{26}
   \and   L.~Lanteri                  \inst{ 1}
   \and   J.-F.~Le~Campion            \inst{19}
   \and   C.-U.~Lee                   \inst{28}
   \and   P.~Leto                     \inst{29}
   \and   H.-C.~Lin                   \inst{30}
   \and   E.~Lindfors                 \inst{ 6}
   \and   M.~G.~Mingaliev             \inst{31}
   \and   S.~Mizoguchi                \inst{13}
   \and   F.~Nicastro                 \inst{21}
   \and   M.~G.~Nikolashvili          \inst{ 8}
   \and   S.~Nishiyama                \inst{13}
   \and   L.~\"Ostman                 \inst{32}
   \and   E.~Ovcharov                 \inst{33}
   \and   P.~P\"a\"akk\"onen          \inst{34}
   \and   M.~Pasanen                  \inst{ 6}
   \and   E.~Pian                     \inst{35}
   \and   T.~Rector                   \inst{36}
   \and   J.~A.~Ros                   \inst{24}
   \and   K.~Sadakane                 \inst{13}
   \and   J.~H.~Selj                  \inst{37}
   \and   E.~Semkov                   \inst{38}
   \and   D.~Sharapov                 \inst{ 4}
   \and   A.~Somero                   \inst{ 4,39}
   \and   I.~Stanev                   \inst{33}
   \and   A.~Strigachev               \inst{38}
   \and   L.~Takalo                   \inst{ 6}
   \and   K.~Tanaka                   \inst{13}
   \and   M.~Tavani                   \inst{40}
   \and   I.~Torniainen               \inst{26}
   \and   M.~Tornikoski               \inst{26}
   \and   C.~Trigilio                 \inst{17}
   \and   G.~Umana                    \inst{17}
   \and   S.~Vercellone               \inst{41}
   \and   A.~Valcheva                 \inst{33,38}
   \and   L.~Volvach                  \inst{14}
   \and   M.~Yamanaka                 \inst{13}
 }

   \offprints{C. M. Raiteri}

   \institute{
 % 1
          INAF, Osservatorio Astronomico di Torino, Italy                                                     
 %           2
   \and   Astron.\ Inst., St.-Petersburg State Univ., Russia                                                  
 %           3
   \and   Pulkovo Observatory, St.\ Petersburg, Russia                                                        
 %           4
   \and   Nordic Optical Telescope, Santa Cruz de La Palma, Spain                                             
 %           5
   \and   Ulugh Beg Astron.\ Inst., Academy of Sciences of Uzbekistan, Tashkent, Uzbekistan                   
 %           6
   \and   Tuorla Observatory, Univ.\ of Turku, Piikki\"{o}, Finland                                           
 %           7
   \and   Department of Astronomy, University of Michigan, MI, USA                                            
 %           8
   \and   Abastumani Astrophysical Observatory, Georgia                                                       
 %           9
   \and   INAF, IASF-Bologna, Italy                                                                           
 %          10
   \and   Michael Adrian Observatory, Trebur, Germany                                                         
 %          11
   \and   IESL, FORTH, Heraklion, Crete, Greece                                                               
 %          12
   \and   Physics Department, University of Crete, Greece                                                     
 %          13
   \and   Astronomical Institute, Osaka Kyoiku University, Japan                                              
 %          14
   \and   Radio Astronomy Lab.\ of Crimean Astrophysical Observatory, Ukraine                                 
 %          15
   \and   Max-Planck-Institut f\"ur Radioastronomie, Bonn, Germany                                            
 %          16
   \and   Department of Physics and Astronomy, Ohio Univ., OH, USA                                            
 %          17
   \and   INAF, Osservatorio Astrofisico di Catania, Italy                                                    
 %          18
   \and   Osservatorio Astronomico della Regione Autonoma Valle d'Aosta, Italy                                
 %          19
   \and   Universit\'e Bordeaux 1/OASU -- CNRS/UMR 5804, France                                               
 %          20
   \and   Instituto de Astrof\'isica de Andaluc\'ia, Granada, Spain                                           
 %          21
   \and   INAF, Osservatorio Astronomico di Roma, Italy                                                       
 %          22
   \and   INAF, Osservatorio Astronomico di Collurania Teramo, Italy                                          
 %          23
   \and   Center for Astrophysics, Guangzhou University, China                                                
 %          24
   \and   Agrupaci\'o Astron\`omica de Sabadell, Spain                                                        
 %          25
   \and   YNAO, Chinese Academy of Sciences, Kunming, China                                                   
 %          26
   \and   Mets\"ahovi Radio Obs., Helsinki Univ.\ of Technology, Finland                                      
 %          27
   \and   Astro Space Center of Lebedev Physical Inst., Moscow, Russia                                        
 %          28
   \and   Korea Astronomy and Space Science Institute, South Korea                                            
 %          29
   \and   INAF, Istituto di Radioastronomia, Sezione di Noto, Italy                                           
 %          30
   \and   Institute of Astronomy, National Central University, Taiwan                                         
 %          31
   \and   Special Astrophysical Observatory, Russia                                                           
 %          32
   \and   Dept.\ of Physics, Stockholm University, Sweden                                                     
 %          33
   \and   Sofia University, Bulgaria                                                                          
 %          34
   \and   Univ.\ of Joensuu, Dept.\ of Physics and Mathematics, Finland                                       
 %          35
   \and   INAF, Osservatorio Astronomico di Trieste, Italy                                                    
 %          36
   \and   University of Alaska Anchorage, AK,  USA                                                            
 %          37
   \and   Inst.\ of Theoretical Astrophysics, Univ.\ of Oslo, Norway                                          
 %          38
   \and   Inst.\ of Astronomy, Bulgarian Academy of Sciences, Sofia, Bulgaria                                 
 %          39
   \and   Observatory, Univ.\ of Helsinki, Finland                                                            
 %          40
   \and   INAF, IASF-Roma, Italy                                                                              
 %          41
   \and   INAF, IASF-Milano, Italy                                                                            
 }

   \date{}

% \abstract{}{}{}{}{} 
% 5 {} token are mandatory
 
  \abstract
   {The quasar-type blazar \object{3C 454.3} was observed to undergo an unprecedented optical outburst in spring 2005, 
affecting the source brightness from the near-IR to the X-ray frequencies.
This was first followed by a millimetric and then by a radio outburst, 
which peaked in February 2006.}
   {In this paper we report on follow-up observations to study the multiwavelength 
emission in the post-outburst phase.}
   {Radio, near-infrared, and optical monitoring was performed by the Whole Earth Blazar Telescope (WEBT)
collaboration in the 2006--2007 observing season. 
XMM-Newton observations on July 2--3 and December 18--19, 2006 
added information on the X-ray and UV states of the source.}
   {The source was in a faint state. 
The radio flux at the higher frequencies showed a fast decreasing trend, which represents the tail of the big radio outburst. 
It was followed by a quiescent state, common at all radio frequencies. 
In contrast, moderate activity characterized the near-IR and optical light curves, with
a progressive increase of the variability amplitude with increasing wavelength.
We ascribe this redder-when-brighter behaviour to the presence of a ``little blue bump" due to line emission 
from the broad line region, which is clearly visible in the source spectral energy distribution (SED) during faint states.
Moreover, the data from the XMM-Newton Optical Monitor reveal a rise of the SED in the ultraviolet, suggesting
the existence of a ``big blue bump" due to thermal emission from the accretion disc.
The X-ray spectra are well fitted with a power-law model with photoelectric absorption, possibly larger than the Galactic one.
However, the comparison with previous X-ray observations would imply that the amount of absorbing matter is variable.
Alternatively, the intrinsic X-ray spectrum presents a curvature, which may depend on the X-ray brightness.
In this case, two scenarios are possible. 
i) There is no extra absorption, and the X-ray spectrum hardens at low energies, the hardening being more evident in bright states.
ii) There is a constant amount of extra absorption, likely in the quasar environment, 
and the X-ray spectrum softens at low energies, at least in faint X-ray states. 
This softening might be the result of a flux contribution by the high-frequency tail of the big blue bump.
}
   {}


   \maketitle
%
%________________________________________________________________

\section{Introduction}

In May 2005 the flat-spectrum radio quasar 3C 454.3 was observed in an unprecedented 
bright optical state. This triggered observations by high-energy satellites 
(Chandra, see \citealt{vil06}; INTEGRAL, see \citealt{pia06}; Swift, see \citealt{gio06}), 
which found an exceptionally high flux also in the X-ray band.
A multiwavelength (radio-to-optical) monitoring campaign was organized by the 
Whole Earth Blazar Telescope (WEBT)\footnote{\tt http://www.to.astro.it/blazars/webt/} to
follow in detail the behaviour of the low-energy emission.
Past data were also collected, both published and unpublished ones, 
so that this behaviour was traced back to summer 1966.
The main results were published by \citet{vil06}: the different behaviour shown by the optical
and radio historical light curves was interpreted as due to the fact that the corresponding 
jet emitting regions are separated and misaligned. In this picture, the inner region, which is responsible 
for the optical radiation, became more aligned with the line of sight during the 2004--2005 outburst.
This produced an increase of the Doppler factor 
and a consequent enhancement of the flux.
Moreover, the analysis of the colour-index behaviour during the outburst, 
generally redder-when-brighter, led \citet{vil06} to suggest the presence of a luminous accretion disc.

The WEBT continued to monitor the source also in the post-outburst period, in particular in 
order to detect a possible correlated event in the radio bands. 
Indeed, a huge mm outburst was observed to peak in June--July 2005.
At the high radio frequencies (43 to $\sim 22$ GHz), a long-lasting outburst developed, reaching the
maximum flux levels in late February 2006. The event was seen progressively delayed and fainter 
going towards lower frequencies, disappearing below 8 GHz.
These data were presented and discussed by \citet{vil07}. According to their interpretation,
the radio peak observed in late February 2006 is not the delayed radio counterpart of the 
spring 2005 optical peak, but it is instead connected to a minor optical flare detected in 
October--November 2005. This interpretation combines an intrinsic variability mechanism
(disturbances travelling down the jet) with a differential change of the
emitting regions viewing angles, due to the motion of the curved jet.

Monitoring of the source by the WEBT continued and was complemented, in July and December 2006,
by two pointings of the XMM-Newton satellite to study the high-energy emission in the post-outburst phase.
In this paper we present the results of this new observing effort on 3C 454.3. 
A third XMM-Newton pointing was performed in May 2007 and its results
will be reported in a further paper, where the multifrequency historical behaviour of the source will
be reconstructed and analysed.

This paper is organised as follows:
the radio-to-optical observations by the WEBT are presented in Sect.\ 2, while Sect.\ 3 
reports on the results of the XMM-Newton pointings. The broad-band spectral energy distributions (SEDs) 
of the source at various epochs are analysed in Sect.\ 4. 
Conclusions are drawn in Sect.\ 5.

\begin{table}
\caption{Ground-based observatories participating in this work.}
\label{obs}
\centering
\begin{tabular}{l r c  }
\hline\hline
Observatory    & Tel.\ size     & Bands\\%               & $N_{\rm obs}$\\
\hline
\multicolumn{3}{c}{\it Radio}\\
\hline
SAO RAS (RATAN-600), Russia  & 600 m$^a$ & 1, 2.3, 5, 8, \\% & 30  \\
                         &               & 11, 22 GHz      \\%&     \\
Crimean (RT-22), Ukraine & 22 m          & 22, 37 GHz          \\%& 174 \\%lat=+44:23:52.6, long=2h 15m 55.1s
Mets\"ahovi, Finland     & 14 m          & 37 GHz              \\%&  62 \\
Noto, Italy              & 32 m          & 43 GHz              \\%&   8 \\
Medicina, Italy          & 32 m          & 5, 8, 22 GHz        \\%&  14 \\
UMRAO, USA               & 26 m          & 5, 8, 14.5 GHz  \\%&  59 \\
\hline
\multicolumn{3}{c}{\it Near-infrared}\\
\hline
Campo Imperatore, Italy  & 110 cm        & $J, H, K$          \\%& 124 \\
\hline
\multicolumn{3}{c}{\it Optical}\\
\hline
Osaka Kyoiku, Japan      &  51 cm        & $V, R, I$          \\%& 416/1078 (48)\\
Yunnan, China            & 102 cm        & $V, R$             \\%& 91/106 (6)\\
Sobaeksan, South Korea   &  61 cm        & $R$                \\%& 102/103 (2)\\%Long=+128 27 27.36, Lat=+36 56 03.89, Alt=1378 m
Lulin (SLT), Taiwan      &  40 cm        & $V, R$             \\%& 2/2 (1)\\
Mt. Maidanak (AZT-22), Uzbekistan & 150 cm        & $U, B, V, R, I$    \\%& /7 (1)\\%Long=+66.8964 deg, Lat=+38.6733 deg (for 1.5m telescope), Alt=2593 m
Mt. Maidanak (T60-K), Uzbekistan &  60 cm        & $U, B, V, R, I$    \\%& 608+4/1136+10 (91+1)\\%Long=+66.8964 deg, Lat=+38.6733 deg (for 1.5m telescope), Alt=2593 m
Abastumani, Georgia      &  70 cm        & $R$                \\%& 660/782 (19)\\%Long=+34.0125 deg, Lat=+44.7266 deg, Alt=650 m
Crimean, Ukraine         &  70 cm        & $B, V, R, I$       \\%& 261/289 (53)\\
Jakokoski, Finland       &  50 cm        & $R$                \\%& 1/1 (1)\\%Long=+29.9969 deg, Lat=+62.7275 deg, Alt=155 m
Skinakas, Greece         & 130 cm        & $B, V, R, I$       \\%& 148/148 (39)\\%Long=+24.897 deg, Lat=+35.212 deg, Alt=1750 m
Rozhen, Bulgaria         & 200 cm        & $U, B, V, R$       \\%& 4/5 (1)\\%Long=+24.7439 deg, Lat=+41.6931, Alt=1759 m
Rozhen, Bulgaria         & 50/70 cm      & $U, B, V, R, I$    \\%& 146/165 (7)\\%Long=+24.7439 deg, Lat=+41.6931, Alt=1759 m
Tuorla, Finland          & 103 cm        & $R$                \\%& 6/7 (2)\\%Long=+22.17 deg, Lat=+60.27.
Michael Adrian, Germany  & 120 cm        & $R$                \\%& 180/212 (16)\\%Long=+8.4114 deg, Lat=+49.9254, Alt=103 m
Valle d'Aosta, Italy     &  81 cm        & $B, V, R, I$       \\%& 40/61 (10)\\%Long=+7.47833 deg, Lat=+45.7895 deg, Alt=1600 m
Sabadell, Spain          &  50 cm        & $R$                \\%& 15/? (11)\\
L'Ampolla, Spain         &  36 cm        & $V, R$             \\%& 7/7 (6)\\%Long=+0.670136 deg, Lat=+40.807 deg, Alt=75 m
Bordeaux, France         &  20 cm        & $V$                \\%& 1/1 (1)\\
Roque (KVA), Spain       &  35 cm        & $R$                \\%& 259/346 (60)\\
Roque (NOT), Spain       & 256 cm        & $U, B, V, R, I$    \\%& 70/71 (9)\\
Kitt Peak (WIYN), USA    &  90 cm        & $B, V, R, I$       \\%& 4/4 (1)\\
Kitt Peak (MDM), USA     & 130 cm        & $U, B, V, R, I$    \\%& 151/234 (4)\\
\hline
\multicolumn{3}{l}{$^a$ Ring telescope}
\end{tabular}
\end{table}

\begin{figure*}
\sidecaption
\includegraphics[width=14cm]{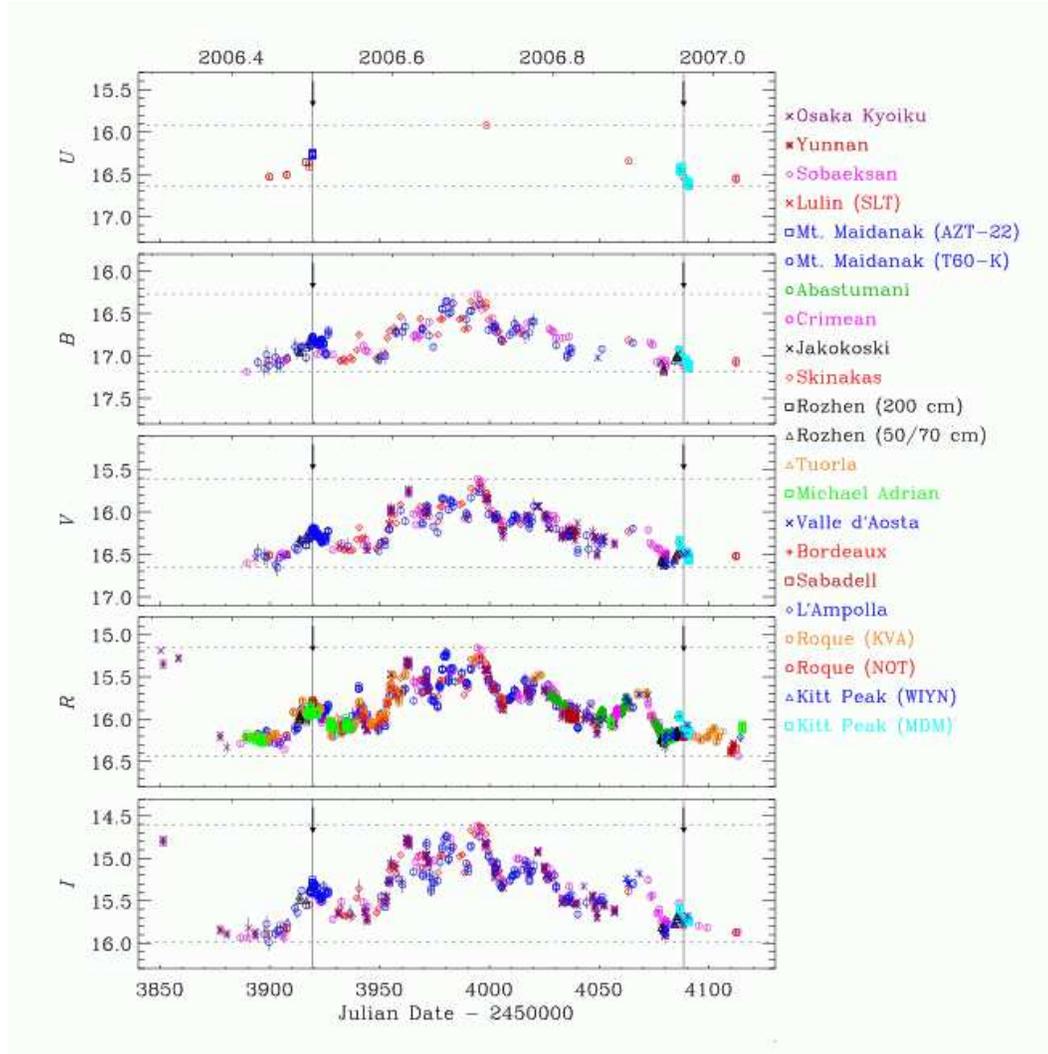}
\caption{Johnson-Cousins $UBVRI$ light curves of 3C 454.3 from May 2006 to January 2007.
The vertical lines and arrows indicate the times of the XMM-Newton pointings of July and December 2006.
Horizontal dotted lines mark the minimum and maximum brightness levels.}
\label{ubvri}
\end{figure*}

\section{Observations by the WEBT}

   \begin{figure}
   \centering
   \resizebox{\hsize}{!}{\includegraphics{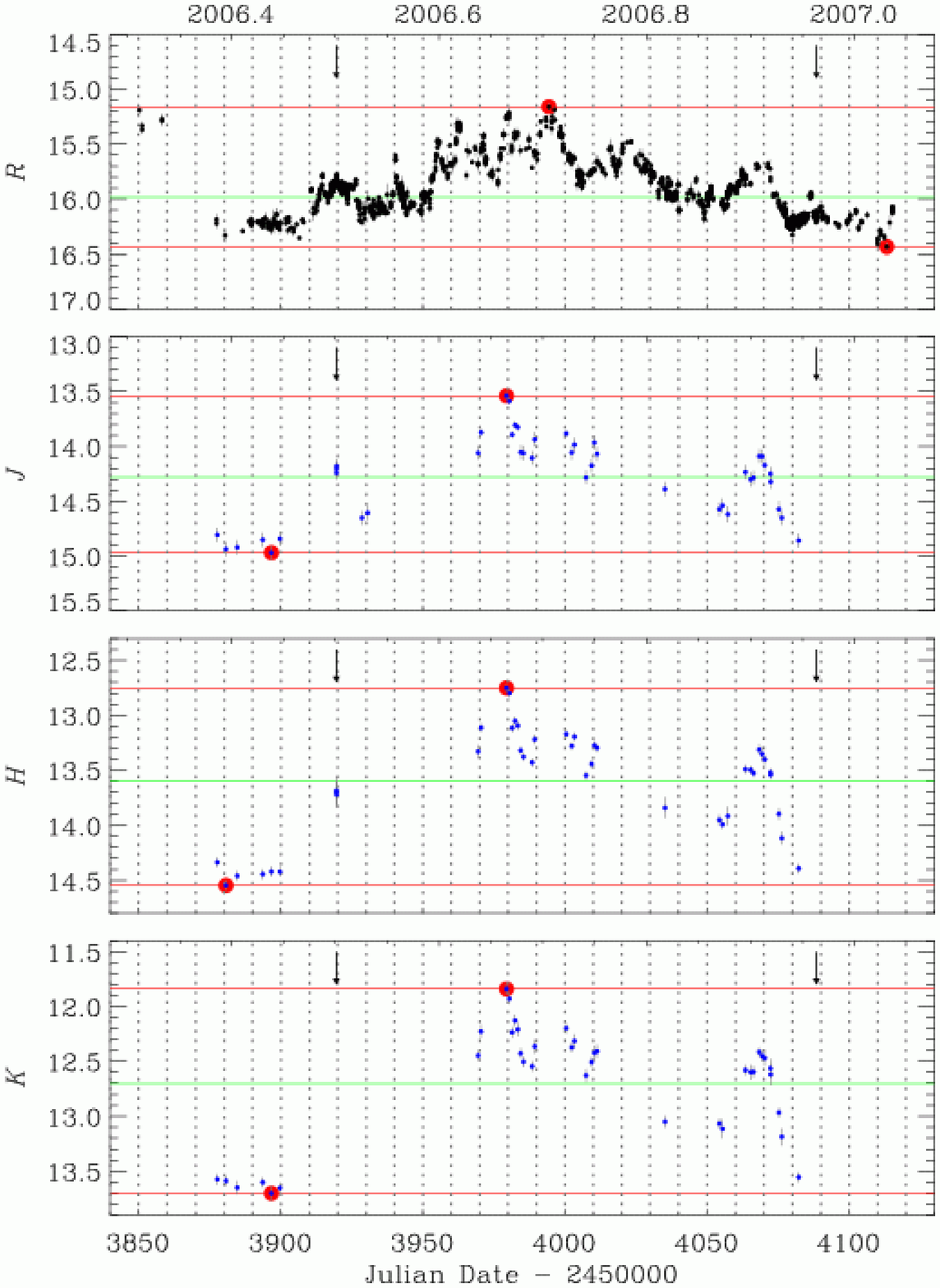}}
      \caption{$R$-band (top panel) and $JHK$ light curves of
3C 454.3 from May 2006 to January 2007. All near-infrared data are from Campo Imperatore.
Maximum and minimum (red lines) as well as average (green lines) brightness levels are indicated.
Arrows mark the times of the XMM-Newton pointings of July and December 2006.} 
         \label{rjhk}
   \end{figure}

Table~\ref{obs} contains the list of the radio, near-infrared, and optical observatories
participating in the 2006--2007 WEBT campaign (starting from September 2006 for the radio observers, 
and from May 2006 for the near-infrared and optical ones). 
Earlier data were published by \citet{vil06,vil07}.
Col.~1 reports the name of the observatory and the country where it is located,
Col.~2 gives the telescope size, and Col.~3 the observing bands. 
Notice that for each group (radio, near-infrared, and optical), the observatories are
listed in order of longitude; indeed, one of the WEBT characteristics is the spread
in longitude of its members, which in principle allows continuous 24 hour monitoring.

Light curves in Johnson-Cousins $UBVRI$ bands are plotted in Fig.~\ref{ubvri}.
The largest contributions (more than 30 observing nights)
came from the Mt.\ Maidanak, Roque (KVA), Crimean, Osaka Kyoiku, and Skinakas observatories.
The source magnitude has been calibrated according to \citet{ang71} in the $U$ band, to \citet{rai98}
in the $BVR$ bands, and to \citet{gon01} in the $I$ band. A cleaning process was applied to minimize
data scattering due to photometric uncertainties, as described by e.g.\ \citet{vil02} and \citet{rai05}.
The total number of data points in Fig.~\ref{ubvri} is 3201, $\sim 63$\% of which are in the $R$ band.

Figure~\ref{ubvri} shows the source in a rather faint state, but with significant magnitude variations:
the difference between the minimum and maximum brightness levels is 0.72, 0.91, 1.04, 1.28, and 1.38 mag in the 
$U$, $B$, $V$, $R$, and $I$ band, respectively. Notwithstanding the different sampling of the light curves,
a progressive increasing of the variability amplitude with wavelength is clearly recognizable.

The $R$-band light curve is compared to the near-IR ones in Fig.~\ref{rjhk}.
The latter are less sampled than the optical ones, since only the 110 cm telescope at Campo Imperatore was
monitoring the source at these frequencies. 
The $JHK$ fluxes have a maximum at JD = 2453979, in correspondence to one of the brightest optical peaks. 
We can see that the variability amplitude continues to increase with wavelength 
(1.43, 1.80, and 1.86 mag in $J$, $H$, and $K$ bands, respectively). 
This trend seems to be another indication in favour of the existence of a luminous accretion disc,
which was suggested to be responsible for the
redder-when-brighter behaviour found by \citet{vil06}.
We will come back to this point in the following section.

\begin{figure*}
\sidecaption
\includegraphics[width=14cm]{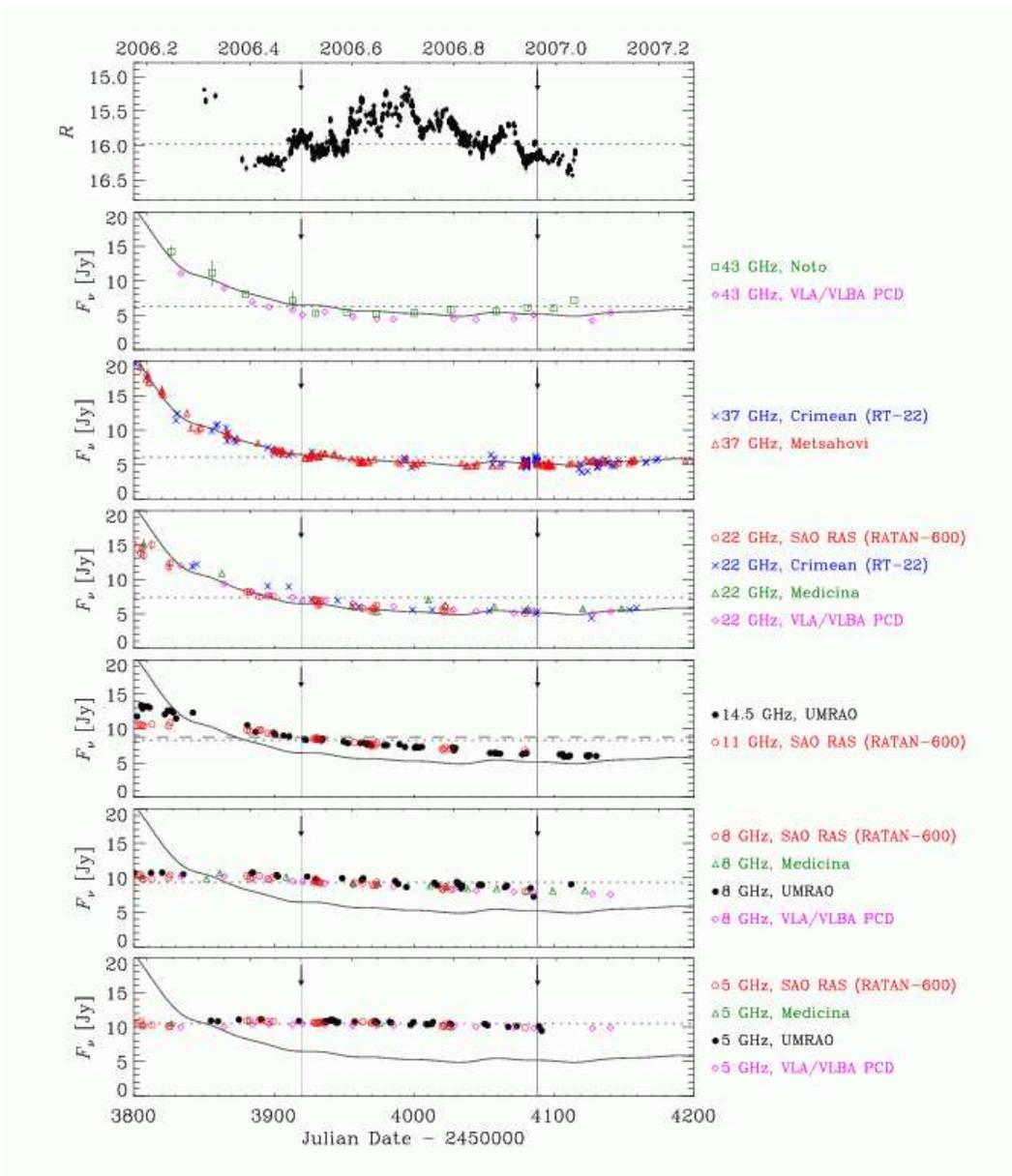}
\caption{$R$-band light curve (top panel) and radio light curves at different frequencies of
3C 454.3 from March 2006 to April 2007.
The vertical lines and arrows indicate the times of the XMM-Newton pointings of July and December 2006.
The dotted (dashed in the case of the 11 GHz light curve) horizontal lines show the average brightness levels
in the considered period. The grey, solid curve represents a cubic spline interpolation through the 20-day binned 37 GHz data.}
\label{radop}
\end{figure*}

The behaviour of the radio flux density (Jy) at different frequencies is shown in Fig.~\ref{radop}, 
where the first panel reports the optical light curve in the $R$ band for a comparison.
We also included data from the VLA/VLBA Polarization Calibration 
Database (PCD)\footnote{\tt http://www.vla.nrao.edu/astro/calib/polar/}.

In contrast with the optical light curves, showing some activity, the radio flux displays only a smooth decreasing trend, 
which is mainly recognizable at the higher frequencies, 
where we see the tail of the big radio outburst peaking in late February 2006 that was analysed by \citet{vil07}.
Indeed, we notice that at the beginning of the period considered in Fig.~\ref{radop}, the radio spectrum is still inverted,
as during the outburst, suggesting that the flux enhancement comes from the inner radio emitting region.
Then, as the high-frequency flux decays, the radio spectrum becomes softer and softer, and 
the flux density increases with increasing wavelength. 
This behaviour is highlighted in Fig.~\ref{radop} by reporting the 37 GHz (20-day binned) cubic
spline interpolation in the various radio panels for a comparison between frequencies.
As expected, the 37 GHz spline matches the 43 GHz data fairly well, while lower-frequencies light curves
intersect the spline at some time, when the corresponding spectral index changes sign and the spectrum is no longer inverted.

\section{Observations by XMM-Newton}

The X-ray Multi-Mirror Mission (XMM) - Newton satellite observed 3C 454.3
twice during the period considered in this paper (PI: C.\ M.\ Raiteri).
The first time was during revolution number 1202, from July 2, 2006 at 21:25:07 UT
to July 3 at 01:58:37 UT (JD = 2453919.39244--2453919.58237).
The second observation took place during revolution number 1287,
from December 18, 2006 at 20:07:27 UT to December 19 at 00:25:14 
(JD = 2454088.33851--2454088.51752).

\subsection{Results from EPIC}

The European Photon Imaging Camera (EPIC) includes three detectors: 
MOS1, MOS2 \citep{tur01}, and pn \citep{str01}.
Since a bright state of the source could not be excluded, 
a medium-filter/small-window configuration
was chosen in order to avoid possible contamination
by lower-energy photons as well as photon pile-up.

Data were reduced with the Science Analysis System (SAS) software, version 7.0.
Only the good time intervals were selected, 
i.e.\ the periods which are free of high-background flares.
This temporal filtering, which was performed according to standard prescriptions, 
reduced the available integration time for MOS1, MOS2, and pn 
by $\sim 4$\%, 5\%, and 30\%, respectively in July.
For the December data these numbers became $\sim 25$\%, 27\%, and 40\%, because of a very high background 
at the beginning of the exposure.

Both the source and background spectra were extracted by setting
{\tt (FLAG==0)} and {\tt (PATTERN$<$=4)} in the selection expression for all the three EPIC detectors.
The first string rejects artifacts as well as events next to both CCD edges and bad pixels,
which may have incorrect energies; the second string selects only single and double pixel events, 
which have the best energy calibration. 
Source spectra were extracted from circular regions with $\sim 35$ and $\sim 40$ arcsec radii for MOS and pn, respectively;
background spectra were selected as the largest source-free circles that can be arranged on the same CCD: 
$\sim 20$ and $\sim 40$ arcsec radius regions for MOS and pn, respectively.

By means of the {\tt grppha} task of the FTOOL package, the source spectra were grouped, i.e.\ the energy channels were binned in order to have a minimum of 25 counts in each bin, and they were associated with the corresponding background and response files.
The grouped spectra were then analysed with the {\tt Xspec} package, version 11.3.2.
Only energy channels between 0.3 and 12 keV were considered.

The same model spectrum was applied to the MOS1, MOS2, and pn data simultaneously to increase the statistics.
We first applied a single power law with Galactic absorption modelled according to the \citet{wil00}
prescriptions and $N_{\rm H}=0.724 \times 10^{21} \, \rm cm^{-2}$, from the Leiden/Argentine/Bonn (LAB) Survey \citep[see][]{kal05}.
The results are shown in Table~\ref{pow}, where 
Col.~2 reports the column density,
Col.~3 the photon spectral index $\Gamma$,
Col.~4 the unabsorbed flux density at 1 keV,
Col.~5 the 2--10 keV observed flux,
and Col.~6 the value of $\chi^2/\nu$ (with the number of degrees of freedom $\nu$).

\begin{table*}
\caption{Results of fitting the EPIC data with different models}             
\label{pow}      
\centering  
\begin{tabular}{ c c c c c c}  
\hline\hline            
Date & $N_{\rm H}$ & $\Gamma$ & $F_{\rm 1 keV}$ & $F_{\rm 2-10 \, keV}$  & $\chi^2/\nu$ ($\nu$) \\
     & [$10^{21} \, \rm cm^{-2}$]&   & [$\mu$Jy]       & [$\rm erg \, cm^{-2} \, s^{-1}$]  &   \\     
\hline                         
\multicolumn{6}{c}{Single power law with Galactic absorption}\\
July 2--3    & 0.724 & 1.52 $\pm$ 0.01 & 0.87 $\pm$ 0.01 & $7.09 \times 10^{-12}$ & 0.857 (1093)\\      
Dec.\ 18--19 & 0.724 & 1.57 $\pm$ 0.01 & 1.18 $\pm$ 0.01 & $8.86 \times 10^{-12}$ & 1.051 (1074)\\
\hline                                   
\multicolumn{6}{c}{Single power law with free absorption}\\
July 2--3    & 0.87 $\pm$ 0.06 & 1.55 $\pm$ 0.02 & 0.90 $\pm$ 0.02 & $6.96 \times 10^{-12}$ & 0.842 (1092)\\      
Dec.\ 18--19 & 1.01 $\pm$ 0.06 & 1.65 $\pm$ 0.02 & 1.29 $\pm$ 0.02 & $8.53 \times 10^{-12}$ & 0.985 (1073)\\
\hline
\multicolumn{6}{c}{Double power law with fixed extra absorption}\\
July 2--3    & 1.34  & 1.49, 2.99 $\pm$ 0.20 & 0.99 $\pm$ 0.02 & $7.03 \times 10^{-12}$ & 0.849 (1092)\\      
Dec.\ 18--19 & 1.34  & 1.58, 2.87 $\pm$ 0.23 & 1.37 $\pm$ 0.04 & $8.60 \times 10^{-12}$ & 0.984 (1073)\\
\hline
\end{tabular}
\end{table*}

\begin{figure}
   \centering
   \resizebox{\hsize}{!}{\includegraphics[angle=-90]{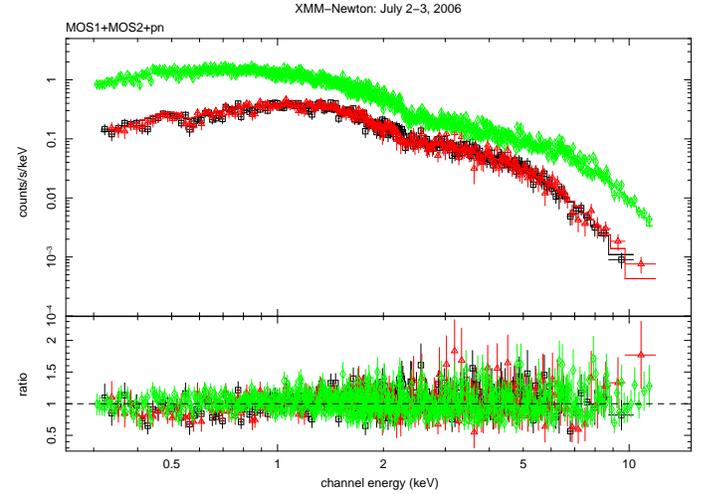}}
      \caption{EPIC spectrum of 3C 454.3 on July 2--3, 2006; 
black squares, red triangles, and green diamonds represent MOS1, MOS2, and pn data, respectively.
The bottom panel shows the ratio between the data and the folded model, a power law with Galactic absorption.} 
         \label{xmm1}
   \end{figure}

\begin{figure}
   \centering
   \resizebox{\hsize}{!}{\includegraphics[angle=-90]{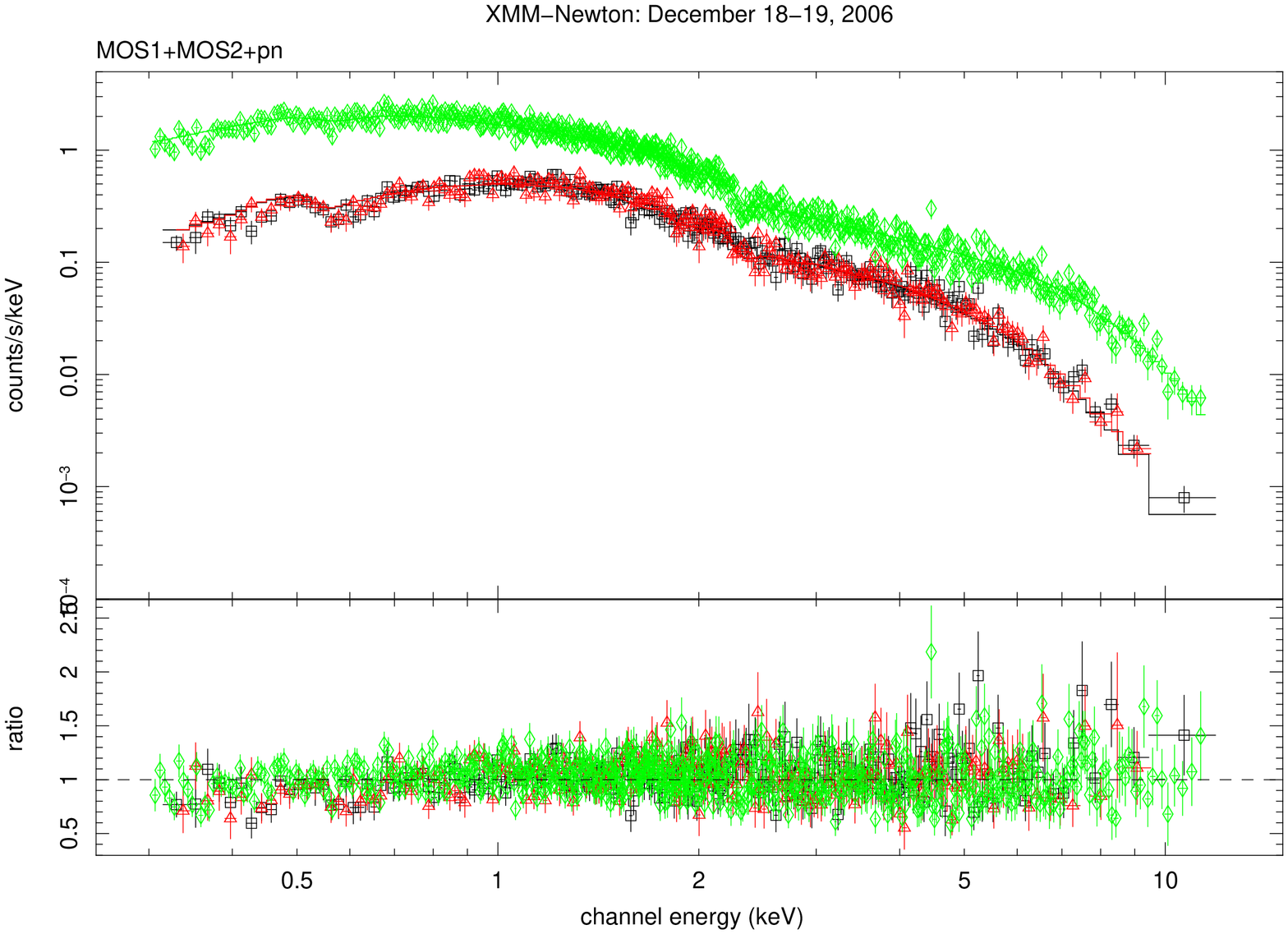}}
      \caption{EPIC spectrum of 3C 454.3 on December 18--19, 2006; 
black squares, red triangles, and green diamonds represent MOS1, MOS2, and pn data, respectively.
The bottom panel shows the ratio between the data and the folded model, a power law with Galactic absorption.} 
         \label{xmm2}
   \end{figure}

The three EPIC spectra fitted with the above model are displayed in the top panels of 
Fig.~\ref{xmm1} (July 2--3) and Fig.~\ref{xmm2} (December 18--19), while
the bottom panels show the ratio between the data and the folded model. 
Assuming a flat cosmology with $H_0=71 \, \rm km \, s^{-1} \, Mpc^{-1}$ and $\Omega_{\rm M}=0.27$, 
the luminosity of the source in the 2--10 keV rest frame energy range is $1.90 \times 10^{46} \, \rm erg \, s^{-1}$ in July, and 
$2.46 \times 10^{46} \, \rm erg \, s^{-1}$ in December.

We notice that fitting previous X-ray spectra of 3C 454.3 often required extra absorption.
\citet{vil06} reported a value
 of $N_{\rm H}=(1.34 \pm 0.05) \times 10^{21} \, \rm cm^{-2}$ 
for the Chandra observation of May 2005, during the outburst phase\footnote{Notice that 
there was a misprint in the \citet{vil06} paper, because of which 1.34 became 13.4.}. 
In that case the unabsorbed 1 keV flux density was $\sim 14$ times higher, but the power-law slope was
very similar: $\Gamma = 1.477 \, \pm \, 0.017$.
An even higher hydrogen column density was found by \citet{gio06} when fitting the 
April--May 2005 data taken by the XRT instrument onboard Swift ($N_{\rm H} \sim 2$--$3 \times 10^{21} \, \rm cm^{-2}$) 
and when reanalysing the BeppoSAX data of June 2000. 

By looking at Fig.\ \ref{xmm2}, we see that the model slightly overestimates the data at the low-energy end
of the plot. This could be an indication that some extra absorption is needed also in this case.
To check this point, we reanalysed the XMM-Newton data letting  $N_{\rm H}$ to vary freely, and obtained
$N_{\rm H}=(0.87 \pm 0.06) \times 10^{21} \, \rm cm^{-2}$ for July, 
and $(1.01 \pm 0.06) \times 10^{21} \, \rm cm^{-2}$ for December (see Table~\ref{pow}).
The F-test probability in the two cases is $5.2 \times 10^{-6}$ and $4.6 \times 10^{-17}$, respectively, indicating
that from a statistical point of view the hypothesis of extra absorption is reasonable.

Another possibility is that the need of extra absorption hides the fact that the intrinsic spectrum 
of the source is not a power law, but presents some hardening in the soft X-ray range \citep{tav07}. 
We investigated this hypothesis by 
fitting the XMM-Newton data with both a broken and a double power-law model with Galactic absorption, 
but the results do not imply any significant curvature of the source spectrum,
and do not improve the goodness of fit. 
However, we cannot rule out that this scenario holds true in case of high X-ray states, like those observed in spring 2005, which seem to require more extra absorption.

On the other hand, the hypothesis that the amount of absorbing material is variable, even
on relatively short time scales, is difficult to explain.
It may be that extra absorption is indeed present, and it is as large as found by e.g.\ \citet{vil06}, 
so that the intrinsic spectrum softens at low X-ray frequencies during faint X-ray states.
We checked this point by fitting a double power-law model with $N_{\rm H}=1.34 \times 10^{21} \, \rm cm^{-2}$
to both the July and December data.
In order to reduce the uncertainties, we also fixed one of the two spectral indices by fitting the data 
above 2 keV with a single power law, since absorption plays a negligible role at these energies.
The results of the fits, which appear to be as good as in the power-law 
with free absorption case (the F-test probability now being $7.5 \times 10^{-4}$ and $2.3 \times 10^{-17}$ 
for the July and December epochs, respectively), are shown in Table~\ref{pow}.

\subsection{Results from OM}

Besides the X-ray detectors, XMM-Newton also carries a co-aligned 30 cm
optical--UV telescope \citep{mas01}, the Optical Monitor (OM). The instrument is equipped with
optical $VBU$ filters, ultraviolet UV$W1$, UV$M2$, UV$W2$ filters, as well as optical and UV grisms.
In both the July and December pointings, we chose to use a $B$, $U$, UV$W1$, UV$M2$ sequence,
as a compromise between a good spectral coverage and the limited duration of the observation.
Exposure times for each filter are given in Cols.~5 and 8 of Table~\ref{om}
for the July and December observations, respectively.

The OM data were reduced with the {\tt omichain} task of SAS version 7.0, and the results were analysed with
{\tt omsource}. Source magnitudes are reported in Cols.~6 and 9 of Table~\ref{om}.
The error on the source magnitude also takes into account 
the dispersion of the results obtained by varying the parameters of the aperture photometry, in particular the 
location and size of the regions from which the background is extracted.

We noticed that the $U$ and $B$ magnitudes of the brightest stars in the field are in fair
agreement with the ground-based calibrations in the $U$ band by \citet{ang71} 
and in the $B$ band by \citet{rai98}, which we have adopted for the WEBT data (see Sect.\ 2).
The UV$W1$ magnitudes of Stars 1 and 4 in the \citet{rai98} notation are $16.04 \pm 0.03$ and $16.52 \pm 0.04$, respectively. In the UV$M2$ frames only the source is measurable.

We derived the Galactic extinction (mag) in the various bands (Col.~2 of Table~\ref{om})
by adopting the $B$-band value of \citet{sch98} and then applying the equations by \citet{car89}.
Extinction in the UV$M2$ band is almost 1 mag; had we adopted the most recent 
prescriptions by \citet{fit99}, we would have obtained a value which is only $\sim 1.8\%$ lower.

\begin{figure}
   \centering
   \resizebox{\hsize}{!}{\includegraphics{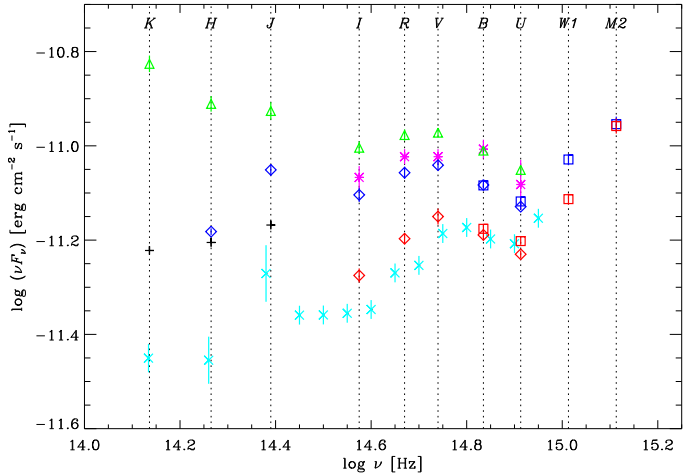}}
      \caption{Spectral energy distributions of 3C 454.3 in the near-IR--UV frequency range (observer's frame).
Blue and red symbols refer to observations performed on July 2--3 and December 18--19, 2006, respectively:
squares represent $B \, U \, W1 \, M2$ data taken by the Optical Monitor; 
diamonds show simultaneous data taken with the NOT in the $UBVRI$ bands and at Campo Imperatore
in the $JH$ bands. Green triangles display the September 2005 SED reported by \citet{vil06}; 
cyan crosses represent observations by \citet{neu79}, black plus signs correspond to the $JHK$ 
data published by \citet{all82}, and pink asterisks are derived from \citet{smi88}.}
   \label{zoom}
   \end{figure}

The transformation of the de-reddened magnitudes into flux densities was obtained by using the method
based on the Vega flux scale\footnote{See {\tt http://xmm.esac.esa.int/sas/7.0.0/watchout/ Evergreen\_tips\_and\_tricks/uvflux.shtml}}:
the adopted Vega magnitudes and flux densities
are reported in Cols.~3 and 4 of Table~\ref{om}, respectively.
The derived 3C 454.3 flux densities corresponding to the July and December pointings are shown 
in Cols.~7 and 10 of Table~\ref{om}, and plotted as blue and red squares in Fig.~\ref{zoom}, respectively.

\begin{table*}
\caption{Results of the OM observations of 3C 454.3.}
\label{om}      
\centering  
\begin{tabular}{ c | c c c | c c c | c c c}  
\hline\hline            
       &            &      &      & \multicolumn{3}{|c|}{July 2--3} & \multicolumn{3}{|c|}{Dec. 18--19}\\
Filter & Extinction & Vega & Vega & $t_{\rm exp}$ &3C 454.3 & 3C 454.3 & $t_{\rm exp}$ &3C 454.3 & 3C 454.3\\
      & [mag]  & [mag]& [$\rm erg \, cm^{-2} \, s^{-1} \, \AA^{-1}$] & [s] &  [mag]   & [mJy] & [s] & [mag] & [mJy]\\     
\hline                        
$B$   & 0.462 & 0.030 & 6.40 $\times 10^{-9}$ & 2900 & 16.82 $\pm$ 0.01 & 1.205 $\pm$ 0.011 & 1402 & 17.05 $\pm$ 0.02 & 0.975 $\pm$ 0.018\\      
$U$   & 0.539 & 0.025 & 3.20 $\times 10^{-9}$ & 2901 & 16.03 $\pm$ 0.02 & 0.931 $\pm$ 0.017 & 1401 & 16.24 $\pm$ 0.02 & 0.767 $\pm$ 0.014\\
$W1$  & 0.648 & 0.025 & 3.68 $\times 10^{-9}$ & 4499 & 15.82 $\pm$ 0.03 & 0.908 $\pm$ 0.025 & 3001 & 16.03 $\pm$ 0.03 & 0.748 $\pm$ 0.021\\
$M2$  & 0.980 & 0.025 & 4.33 $\times 10^{-9}$ & 4600 & 15.89 $\pm$ 0.04 & 0.857 $\pm$ 0.032 & 7520$^a$ & 15.90 $\pm$ 0.04 & 0.849 $\pm$ 0.031\\
\hline                                   
\multicolumn{10}{l}{$^a$ Three subsequent exposures of 3100, 3099, and 1321 s.}
\end{tabular}
\end{table*}

This figure also shows ground-based optical and near-infrared flux densities from this work as well
as from the literature (see description below); all the plotted values have been obtained by correcting 
for the Galactic extinction, using the same method we used for the OM data. 
In order to convert de-reddened magnitudes into flux densities, we adopted the zero-mag fluxes by \citet{bes98}.

Blue and red diamonds in Fig.~\ref{zoom} correspond to the
$UBVRI$ data acquired by the NOT during the July and December XMM-Newton pointings, 
respectively.
As we can see from the figure, the ground-based optical flux densities agree fairly well
with the OM ones in the overlapping frequency range.
Flux densities in the $JH$ bands from observations performed at Campo Imperatore 
simultaneously to the July pointing are also displayed (blue diamonds).

In Fig.~\ref{zoom}, the above two near-infrared--ultraviolet SEDs are compared 
to previous faint-state SEDs.
One SED (green triangles) was obtained by \citet{vil06} from near-infrared and optical observations by the 
WEBT in late September 2005, just after the end of the big outburst.
The faintest-state SED (cyan crosses) was derived from  \citet{neu79}, who
observed this source with the 5 m Hale Telescope at Palomar Mountain; the optical data were 
taken in January 1973, while the near-infrared ones in October 1976.
Further near-infrared data points (black plus signs) 
were derived from the August 1980 observations of \citet{all82}.
Finally, the pink asterisks refer to data acquired in December 1986 by \citet{smi88}.

By looking at Fig.~\ref{zoom} we notice that:
\begin{itemize}
\item all the optical SEDs have a bump shape, with peak in the $V$--$B$ frequency range;
\item the OM data presented in this paper confirm and extend further in frequency the rise of the SED 
in the ultraviolet that was present in the data of \citet{neu79};
\item there is also an upturn from the $I$ to the $J$ band;
\item going towards lower frequencies, the behaviour of the brightest-state SED is different from that
of the lower-state ones, since in the former the SED rise continues, while in all the latter
the values in the $H$ band are lower than in the $J$ one. The difference between the $J$ and $H$ values is 
greater in the July 2006 SED than in the \citet{all82} one, while in the case of \citet{neu79} large
uncertainties affect the data.
\end{itemize} 

The bump peaking around the $V$ and $B$ bands likely corresponds to the {\it little blue bump} 
observed in quasars between $\sim 2000$ and 4000 \AA\ in the rest frame.
This is due to the contribution of many emission lines produced in the broad line region (BLR),
in particular the numerous \ion{Fe}{ii} and the \ion{Mg}{ii} lines, and Balmer continuum \citep{wil85}.
Since the 3C 454.3 redshift is $z=0.859$, 
\ion{Fe}{ii} lines would mostly contribute to the observed spectrum around the $B$ band, 
while the flux in the  $V$ band would be enhanced by the \ion{Mg}{ii} line and Balmer continuum contributions.
In the same way, the flux excess in correspondence of the $J$ band is likely due to a prominent broad H$\alpha$ emission line.
This bump  is more evident when the beamed synchrotron radiation from the jet is fainter. 
The lowest-flux SEDs in Fig.~\ref{zoom} (January 1973 and December 2006) show states where the BLR 
component probably dominates the source emission. In contrast, the steep near-infrared part of the
September 2005 SED suggests that in this epoch the synchrotron component was giving a higher contribution,
and the BLR component, though still recognizable, begins to be diluted by the synchrotron one. 

On the other hand, the rise of the SEDs in the ultraviolet is likely the signature of the {\it big blue bump} observed in many active galactic nuclei, 
which is commonly interpreted as thermal emission from the accretion disc \citep[e.g.][]{lao90}.
Evidences of this thermal component have been found in other quasar-type blazars, 
such as 3C 273 \citep{smi93,von97,gra04,tur06}, 3C 279 \citep{pia99}, and 3C 345 \citep{bre86}. 
Indeed, by separating the polarized (synchrotron) from the unpolarized (thermal)
component in individual quasar IR--UV spectra, \citet{wil91} showed that this latter 
is consistent with the SEDs observed in luminous radio-quiet and lobe-dominated quasars.
In the case of the BL Lacertae object AO 0235+164, a UV--soft-X-ray bump is recognizable in several SEDs, 
but whether this component is thermal radiation from the disc or rather another synchrotron component from 
the jet is not clear yet \citep{rai05,rai06b,rai06a}. 

The fact that the source flux appears to be constant in the UV$M2$ band, which is less affected by the synchrotron contribution, suggests that the big blue bump is a rather non-variable component, at least on a few-month time scale.

The presence of these non-jet components, which mostly affect the blue part of the spectrum,
allows us to understand why the variability amplitude in
the near-IR and optical bands increases with wavelength (i.e.\ the redder-when-brighter behaviour), 
as noticed in Sect.\ 2.
Indeed, when the jet emission decreases, the non-jet contribution sustains the source flux 
more in the blue than in the red.

\section{Broad-band spectral energy distribution}

\begin{figure*}
\sidecaption
\includegraphics[width=12cm]{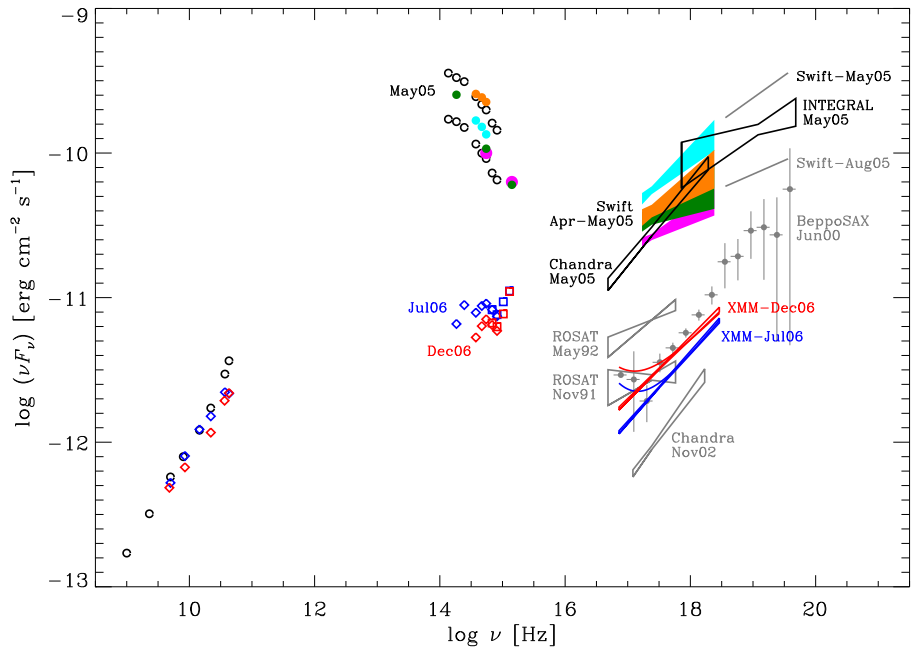}
      \caption{Broad-band spectral energy distribution of 3C 454.3 (observer's frame). 
Flux densities derived from the XMM-Newton (EPIC and OM) observations of July and December 2006 are shown in blue and red, respectively. 
The same colours are used to plot simultaneous low-energy (radio-to-optical) data from observations by the WEBT. 
The X-ray spectra are displayed as the results of spectral fitting by two models: a power law with Galactic absorption (straight lines) and a double power law with fixed extra absorption (curved lines).
X-ray spectra resulting from previous satellite pointings are also shown, in particular those obtained from the Chandra, INTEGRAL, and Swift (XRT and BAT instruments) observations in 2005. 
Swift-XRT spectra are complemented by simultaneous near-IR-to-UV data from the REM telescope and UVOT detector \citep{gio06}, while radio-to-optical data from the WEBT (open circles) indicate brightness levels during the Chandra and INTEGRAL pointings.
}
   \label{sed}
   \end{figure*}

The broad-band SED of 3C 454.3 is shown in Fig.~\ref{sed}.
The near-IR-to-UV SEDs corresponding to the XMM-Newton pointings of July and December 2006, 
which were displayed in Fig.~\ref{zoom} and discussed in the previous section, 
are now complemented by the X-ray spectra presented in Sect.\ 3.1 
(the results of both the power-law with Galactic absorption and 
double power-law with fixed extra absorption models are shown)
and by radio data from 5 to 43 GHz that were
taken at the same time or within 3 days from the XMM-Newton observations. 
The comparison between the UV and X-ray data clearly shows the UV excess. 
As discussed in the previous section, this excess is the signature of the big blue bump, 
which is most likely due to thermal emission from the accretion disc.
Moreover, fitting the X-ray data with a double power-law with fixed extra absorption model produces also a soft-X-ray excess,
which would indicate a possible contribution from the high-frequency tail of the big blue bump.

We notice that, while the radio-to-optical state was fainter in December than in July 2006, in 
the X-ray domain it was the opposite. However, a linear fit to the high-frequency ($\ge 22$ GHz) radio data in Fig.~\ref{sed} reveals that this part of the radio spectrum was harder in December ($\alpha=0.10 \pm 0.08$, with $F_\nu \propto \nu^{-\alpha}$) than in July {\bf ($\alpha=0.46 \pm 0.03$)}, suggesting a larger flux in the mm bands, whose photons are inverse-Comptonized to the X-ray frequencies we observe with XMM-Newton.

In the same figure we show the results of power-law model fits to X-ray spectra from observations by ROSAT in 1991 and 1992 \citep{sam97,pri96}, 
BeppoSAX in 2000\footnote{The spectrum shown in the figure, corresponding to observations performed in June 2000, 
was derived from the ASI Science Data Center ({\tt http://www.asdc.asi.it/}).},
Chandra in 2002 \citep{mar05}, and from observations performed near the 2005 outburst peak by Chandra \citep{vil06}, INTEGRAL \citep{pia06}, and Swift \citep{gio06}.

The open circles show the radio brightness levels and the near-IR--optical variability range observed by the WEBT members in May 2005, during the INTEGRAL and Chandra pointings, and reported by \citet{vil06}. 
Lower-energy data simultaneous to the 0.7--10 keV observations of the XRT instrument onboard Swift were taken by the UVOT instrument onboard the satellite
(optical--UV) and by the REM telescope (near-IR--optical). They are plotted in the figure distinguishing the four epochs presented by \citet{gio06}: April 24 (pink symbols), May 11 (cyan), May 17 (dark green), and May 19 (orange).
An indication of the hard-X-ray flux (between 15 and 150 keV) registered by the BAT instrument onboard Swift in May and August 2005 is also reported.
The X-ray spectra from Swift (in particular that of May 19) are less hard than the Chandra spectrum acquired on May 19--20; 
this may depend, at least in part, on the 1.7--1.8 times higher $N_{\rm H}$ value adopted by \citet{gio06} with respect to that used by \citet{vil06}.

The data taken in the Swift epochs May 11 and May 19, 2005 reveal the same ``crossing" trend that we have already noticed for the July and December 2006 XMM-Newton epochs, i.e.\ a higher optical flux corresponding to a lower X-ray flux.
However, this does not imply any direct relation between the optical and X-ray emissions, since they come from different regions and, as discussed above for the XMM-Newton epochs, the X-ray emission is most likely correlated to the mm one. 

We notice that the optical--UV data taken during the outburst phase do not show any evidence of 
the blue--UV bumps observed during faint states, as they are overwhelmed by the beamed synchrotron radiation.

\section{Conclusions}

The unprecedented outburst of 3C 454.3 in 2005, affecting the source emission from the near-IR to the X-ray frequencies, 
was first followed by a huge millimetric and then by a long-lasting, extraordinary, high-frequency radio outburst.
After that, the blazar underwent a multifrequency faint-state period from mid 2006 to April 2007, 
which was characterized by moderate variability in the near-infrared and optical bands. 
The variability amplitude was larger at longer wavelengths, consistently with the results of e.g.\ \citet{bre86}
for the quasar-type blazar 3C 345, whereas the general feature of BL Lac-type blazars is 
to show larger-amplitude flux changes at higher frequencies \citep[e.g.][]{vil04a,pap07,wu07}.

After the radio peak of late February 2006, 
at the higher radio frequencies we observed the fast outburst dimming phase, 
with the transition from the inverted radio spectrum, 
which characterised the long-lasting outburst, to the usual, softer one.
According to \citet{vil07}, 
the fast drop at all the higher radio frequencies and the absence of any flux increase at the lower ones
suggest that the event that perturbed the inner radio emitting region has propagated outwards,
in a jet region that is misaligned with respect to the line of sight.
Hence, we see now only the ``quiescent" emission of the source, i.e.\ a higher flux density at larger wavelengths, 
since this radiation comes from outer and more transparent emitting regions.

The faint state allowed us to recognize important spectral features, which are usually hidden by the beamed synchrotron emission from the jet.
The first is the little blue bump in the optical band, peaking around the $V$ and $B$ bands, and likely due to the contribution of \ion{Fe}{ii} and \ion{Mg}{ii} emission lines and Balmer continuum produced in the BLR \citep{wil85}. 
Another minor bump seems to peak in the $J$ band, and this is most likely the signature of a broad and prominent H$\alpha$ line.

But the major feature, which was clearly revealed by the data acquired by the Optical Monitor onboard XMM-Newton, 
is the spectral break in the $U$ band, with the transition from a soft optical spectrum to a hard UV one.
This suggests the presence of a big UV bump most likely due to thermal emission from the accretion disc, 
as already recognized in few other quasar-type blazars, but not yet in 3C 454.3.

These BLR and thermal emission components explain why the optical and near-infrared light curves show an 
increasing variability amplitude when going from the higher to the lower frequencies, as well as the 
redder-when-brighter behaviour noticed by \citet{vil06}. 
Indeed, when the jet emission decreases, the blue part of the optical spectrum cannot go below a certain level 
because of the non-jet contributions.

A power-law model with Galactic photoelectric absorption gives a fair fit to the XMM-Newton X-ray spectra.
Some amount of extra absorption ($\sim 20$\% for the July observation
and $\sim 40$\% for the  December one) yields better results from a statistical point of view. 
Since the need of even much higher values of extra absorption was claimed by various authors when analysing 
past X-ray data, this might suggest the presence of absorbing material with
variable column density, as observed for AO 0235+164 \citep{wol82} and suggested for BL Lacertae \citep{rav03}.

The possibility that the requirement of extra absorption actually hides a hardening
of the intrinsic soft-X-ray spectrum is not supported by our results.
Indeed, applying curved models with Galactic absorption to the XMM-Newton data does not 
produce significant curvatures and does not improve the goodness of fit with respect to the single power-law case.
However, bright X-ray states, which seem to require larger values of extra absorption,
might be charactized by intrinsic curved spectra with hardening in the soft X-ray range.

An alternative explanation is that extra absorption, possibly in the quasar environment, does exist,
and it is not variable. In this case, the fact that X-ray data acquired during faint X-ray states seem
to require a lower amount of absorption could be due to a spectral softening of the lower-energy X-ray spectrum.
This softening might be due to the contribution of the high-frequency tail of the big blue bump.

At present, it is not possible to discriminate among the various models of X-ray spectral fitting.
Further X-ray observations will hopefully help to clarify this matter.

\begin{acknowledgements}
We thank the referee, Beverley J.\ Wills, for useful comments and suggestions.
This work is partly based on observations obtained with XMM-Newton, an ESA science mission
with instruments and contributions directly funded by ESA Member States and NASA, and
on observations made with the Nordic Optical Telescope, operated
on the island of La Palma jointly by Denmark, Finland, Iceland,
Norway, and Sweden, in the Spanish Observatorio del Roque de los
Muchachos of the Instituto de Astrofisica de Canarias.
This research has made use of data from the University of Michigan Radio Astronomy Observatory,
which is supported by the National Science Foundation and by funds from the University of Michigan.
The Torino team acknowledges financial support by the Italian Space Agency through contract ASI/INAF I/023/05/0.
St.\ Petersburg team was supported by  the Russian Fund of Basic Research, grant 05-02-17562.
The Mets\"ahovi team acknowledges the support from the Academy of Finland.
JHF's work is partially supported by the National Natural
Science Foundation of China (10573005,10633010).
YYK is a research fellow of the Alexander von Humboldt Foundation.
\mbox{RATAN--600} observations were partly supported by the
Russian Foundation for Basic Research (project 05-02-17377).
ACG's work is supported by NNSF of China grant no. 10533050.

\end{acknowledgements}


\begin{thebibliography}{41}
\expandafter\ifx\csname natexlab\endcsname\relax\def\natexlab#1{#1}\fi

\bibitem[{{Allen} {et~al.}(1982){Allen}, {Ward}, \& {Hyland}}]{all82}
{Allen}, D.~A., {Ward}, M.~J., \& {Hyland}, A.~R. 1982, \mnras, 199, 969

\bibitem[{{Angione}(1971)}]{ang71}
{Angione}, R.~J. 1971, \aj, 76, 412

\bibitem[{{Bessell} {et~al.}(1998){Bessell}, {Castelli}, \& {Plez}}]{bes98}
{Bessell}, M.~S., {Castelli}, F., \& {Plez}, B. 1998, \aap, 333, 231

\bibitem[{{Bregman} {et~al.}(1986){Bregman}, {Glassgold}, {Huggins},
  {Neugebauer}, {Soifer}, {Matthews}, {Elias}, {Webb}, {Pollock}, {Pica},
  {Leacock}, {Smith}, {Aller}, {Aller}, {Hodge}, {Dent}, {Balonek},
  {Barvainis}, {Roellig}, {Wisniewski}, {Rieke}, {Lebofsky}, {Wills}, {Wills},
  {Ku}, {Bregman}, {Witteborn}, {Lester}, {Impey}, \& {Hackwell}}]{bre86}
{Bregman}, J.~N., {Glassgold}, A.~E., {Huggins}, P.~J., {et~al.} 1986, \apj,
  301, 708

\bibitem[{{Cardelli} {et~al.}(1989){Cardelli}, {Clayton}, \& {Mathis}}]{car89}
{Cardelli}, J.~A., {Clayton}, G.~C., \& {Mathis}, J.~S. 1989, \apj, 345, 245

\bibitem[{{Fitzpatrick}(1999)}]{fit99}
{Fitzpatrick}, E.~L. 1999, \pasp, 111, 63

\bibitem[{{Giommi} {et~al.}(2006){Giommi}, {Blustin}, {Capalbi},
  {Colafrancesco}, {Cucchiara}, {Fuhrmann}, {Krimm}, {Marchili}, {Massaro},
  {Perri}, {Tagliaferri}, {Tosti}, {Tramacere}, {Burrows}, {Chincarini},
  {Falcone}, {Gehrels}, {Kennea}, \& {Sambruna}}]{gio06}
{Giommi}, P., {Blustin}, A.~J., {Capalbi}, M., {et~al.} 2006, \aap, 456, 911

\bibitem[{{Gonz{\'a}lez-P{\'e}rez} {et~al.}(2001){Gonz{\'a}lez-P{\'e}rez},
  {Kidger}, \& {Mart{\'{\i}}n-Luis}}]{gon01}
{Gonz{\'a}lez-P{\'e}rez}, J.~N., {Kidger}, M.~R., \& {Mart{\'{\i}}n-Luis}, F.
  2001, \aj, 122, 2055

\bibitem[{{Grandi} \& {Palumbo}(2004)}]{gra04}
{Grandi}, P. \& {Palumbo}, G.~G.~C. 2004, Science, 306, 998

\bibitem[{{Kalberla} {et~al.}(2005){Kalberla}, {Burton}, {Hartmann}, {Arnal},
  {Bajaja}, {Morras}, \& {P{\"o}ppel}}]{kal05}
{Kalberla}, P.~M.~W., {Burton}, W.~B., {Hartmann}, D., {et~al.} 2005, \aap,
  440, 775

\bibitem[{{Laor}(1990)}]{lao90}
{Laor}, A. 1990, \mnras, 246, 369

\bibitem[{{Marshall} {et~al.}(2005){Marshall}, {Schwartz}, {Lovell}, {Murphy},
  {Worrall}, {Birkinshaw}, {Gelbord}, {Perlman}, \& {Jauncey}}]{mar05}
{Marshall}, H.~L., {Schwartz}, D.~A., {Lovell}, J.~E.~J., {et~al.} 2005, \apjs,
  156, 13

\bibitem[{{Mason} {et~al.}(2001){Mason}, {Breeveld}, {Much}, {Carter},
  {Cordova}, {Cropper}, {Fordham}, {Huckle}, {Ho}, {Kawakami}, {Kennea},
  {Kennedy}, {Mittaz}, {Pandel}, {Priedhorsky}, {Sasseen}, {Shirey}, {Smith},
  \& {Vreux}}]{mas01}
{Mason}, K.~O., {Breeveld}, A., {Much}, R., {et~al.} 2001, \aap, 365, L36

\bibitem[{{Neugebauer} {et~al.}(1979){Neugebauer}, {Oke}, {Becklin}, \&
  {Matthews}}]{neu79}
{Neugebauer}, G., {Oke}, J.~B., {Becklin}, E.~E., \& {Matthews}, K. 1979, \apj,
  230, 79

\bibitem[{{Papadakis} {et~al.}(2007){Papadakis}, {Villata}, \&
  {Raiteri}}]{pap07}
{Papadakis}, I.~E., {Villata}, M., \& {Raiteri}, C.~M. 2007, \aap, 470, 857

\bibitem[{{Pian} {et~al.}(2006){Pian}, {Foschini}, {Beckmann}, {Soldi},
  {T{\"u}rler}, {Gehrels}, {Ghisellini}, {Giommi}, {Maraschi}, {Pursimo},
  {Raiteri}, {Tagliaferri}, {Tornikoski}, {Tosti}, {Treves}, {Villata}, {Barr},
  {Courvoisier}, {di Cocco}, {Hudec}, {Fuhrmann}, {Malaguti}, {Persic},
  {Tavecchio}, \& {Walter}}]{pia06}
{Pian}, E., {Foschini}, L., {Beckmann}, V., {et~al.} 2006, \aap, 449, L21

\bibitem[{{Pian} {et~al.}(1999){Pian}, {Urry}, {Maraschi}, {Madejski},
  {McHardy}, {Koratkar}, {Treves}, {Chiappetti}, {Grandi}, {Hartman}, {Kubo},
  {Leach}, {Pesce}, {Imhoff}, {Thompson}, \& {Wehrle}}]{pia99}
{Pian}, E., {Urry}, C.~M., {Maraschi}, L., {et~al.} 1999, \apj, 521, 112

\bibitem[{{Prieto}(1996)}]{pri96}
{Prieto}, M.~A. 1996, \mnras, 282, 421

\bibitem[{{Raiteri} {et~al.}(2005){Raiteri}, {Villata}, {Ibrahimov},
  {Larionov}, {Kadler}, {Aller}, {Aller}, {Kovalev}, {Lanteri}, {Nilsson},
  {Papadakis}, {Pursimo}, {Romero}, {Ter{\"a}sranta}, {Tornikoski}, {Arkharov},
  {Barnaby}, {Berdyugin}, {B{\"o}ttcher}, {Byckling}, {Carini}, {Carosati},
  {Cellone}, {Ciprini}, {Combi}, {Crapanzano}, {Crowe}, {di Paola}, {Dolci},
  {Fuhrmann}, {Gu}, {Hagen-Thorn}, {Hakala}, {Impellizzeri}, {Jorstad}, {Kerp},
  {Kimeridze}, {Kovalev}, {Kraus}, {Krichbaum}, {Kurtanidze},
  {L{\"a}hteenm{\"a}ki}, {Lindfors}, {Mingaliev}, {Nesci}, {Nikolashvili},
  {Ohlert}, {Orio}, {Ostorero}, {Pasanen}, {Pati}, {Poteet}, {Ros}, {Ros},
  {Shastri}, {Sigua}, {Sillanp{\"a}{\"a}}, {Smith}, {Takalo}, {Tosti},
  {Vasileva}, {Wagner}, {Walters}, {Webb}, {Wills}, {Witzel}, \&
  {Xilouris}}]{rai05}
{Raiteri}, C.~M., {Villata}, M., {Ibrahimov}, M.~A., {et~al.} 2005, \aap, 438,
  39

\bibitem[{{Raiteri} {et~al.}(2006{\natexlab{a}}){Raiteri}, {Villata}, {Kadler},
  {Ibrahimov}, {Kurtanidze}, {Larionov}, {Tornikoski}, {Boltwood}, {Lee},
  {Aller}, {Romero}, {Aller}, {Araudo}, {Arkharov}, {Bach}, {Barnaby},
  {Berdyugin}, {Buemi}, {Carini}, {Carosati}, {Cellone}, {Cool}, {Dolci},
  {Efimova}, {Fuhrmann}, {Hagen-Thorn}, {Holcomb}, {Ilyin}, {Impellizzeri},
  {Ivanidze}, {Kapanadze}, {Kerp}, {Konstantinova}, {Kovalev}, {Kovalev},
  {Kraus}, {Krichbaum}, {L{\"a}hteenm{\"a}ki}, {Lanteri}, {Leto}, {Lindfors},
  {Mattox}, {Napoleone}, {Nikolashvili}, {Nilsson}, {Ohlert}, {Papadakis},
  {Pasanen}, {Poteet}, {Pursimo}, {Ros}, {Sigua}, {Smith}, {Takalo},
  {Trigilio}, {Tr{\"o}ller}, {Umana}, {Ungerechts}, {Walters}, {Witzel}, \&
  {Xilouris}}]{rai06b}
{Raiteri}, C.~M., {Villata}, M., {Kadler}, M., {et~al.} 2006{\natexlab{a}},
  \aap, 459, 731

\bibitem[{{Raiteri} {et~al.}(2006{\natexlab{b}}){Raiteri}, {Villata}, {Kadler},
  {Krichbaum}, {B{\"o}ttcher}, {Fuhrmann}, \& {Orio}}]{rai06a}
{Raiteri}, C.~M., {Villata}, M., {Kadler}, M., {et~al.} 2006{\natexlab{b}},
  \aap, 452, 845

\bibitem[{{Raiteri} {et~al.}(1998){Raiteri}, {Villata}, {Lanteri}, {Cavallone},
  \& {Sobrito}}]{rai98}
{Raiteri}, C.~M., {Villata}, M., {Lanteri}, L., {Cavallone}, M., \& {Sobrito},
  G. 1998, \aaps, 130, 495

\bibitem[{{Ravasio} {et~al.}(2003){Ravasio}, {Tagliaferri}, {Ghisellini},
  {Tavecchio}, {B{\"o}ttcher}, \& {Sikora}}]{rav03}
{Ravasio}, M., {Tagliaferri}, G., {Ghisellini}, G., {et~al.} 2003, \aap, 408,
  479

\bibitem[{{Sambruna}(1997)}]{sam97}
{Sambruna}, R.~M. 1997, \apj, 487, 536

\bibitem[{{Schlegel} {et~al.}(1998){Schlegel}, {Finkbeiner}, \&
  {Davis}}]{sch98}
{Schlegel}, D.~J., {Finkbeiner}, D.~P., \& {Davis}, M. 1998, \apj, 500, 525

\bibitem[{{Smith} {et~al.}(1988){Smith}, {Elston}, {Berriman}, {Allen}, \&
  {Balonek}}]{smi88}
{Smith}, P.~S., {Elston}, R., {Berriman}, G., {Allen}, R.~G., \& {Balonek},
  T.~J. 1988, \apjl, 326, L39

\bibitem[{{Smith} {et~al.}(1993){Smith}, {Schmidt}, \& {Allen}}]{smi93}
{Smith}, P.~S., {Schmidt}, G.~D., \& {Allen}, R.~G. 1993, \apj, 409, 604

\bibitem[{{Str{\"u}der} {et~al.}(2001){Str{\"u}der}, {Briel}, {Dennerl},
  {Hartmann}, {Kendziorra}, {Meidinger}, {Pfeffermann}, {Reppin}, {Aschenbach},
  {Bornemann}, {Br{\"a}uninger}, {Burkert}, {Elender}, {Freyberg}, {Haberl},
  {Hartner}, {Heuschmann}, {Hippmann}, {Kastelic}, {Kemmer}, {Kettenring},
  {Kink}, {Krause}, {M{\"u}ller}, {Oppitz}, {Pietsch}, {Popp}, {Predehl},
  {Read}, {Stephan}, {St{\"o}tter}, {Tr{\"u}mper}, {Holl}, {Kemmer}, {Soltau},
  {St{\"o}tter}, {Weber}, {Weichert}, {von Zanthier}, {Carathanassis}, {Lutz},
  {Richter}, {Solc}, {B{\"o}ttcher}, {Kuster}, {Staubert}, {Abbey}, {Holland},
  {Turner}, {Balasini}, {Bignami}, {La Palombara}, {Villa}, {Buttler},
  {Gianini}, {Lain{\'e}}, {Lumb}, \& {Dhez}}]{str01}
{Str{\"u}der}, L., {Briel}, U., {Dennerl}, K., {et~al.} 2001, \aap, 365, L18

\bibitem[{{Tavecchio} {et~al.}(2007){Tavecchio}, {Maraschi}, {Ghisellini},
  {Kataoka}, {Foschini}, {Sambruna}, \& {Tagliaferri}}]{tav07}
{Tavecchio}, F., {Maraschi}, L., {Ghisellini}, G., {et~al.} 2007, \apj, 665,
  980

\bibitem[{{T{\"u}rler} {et~al.}(2006){T{\"u}rler}, {Chernyakova},
  {Courvoisier}, {Foellmi}, {Aller}, {Aller}, {Kraus}, {Krichbaum},
  {L{\"a}hteenm{\"a}ki}, {Marscher}, {McHardy}, {O'Brien}, {Page}, {Popescu},
  {Robson}, {Tornikoski}, \& {Ungerechts}}]{tur06}
{T{\"u}rler}, M., {Chernyakova}, M., {Courvoisier}, T.~J.-L., {et~al.} 2006,
  \aap, 451, L1

\bibitem[{{Turner} {et~al.}(2001){Turner}, {Abbey}, {Arnaud}, {Balasini},
  {Barbera}, {Belsole}, {Bennie}, {Bernard}, {Bignami}, {Boer}, {Briel},
  {Butler}, {Cara}, {Chabaud}, {Cole}, {Collura}, {Conte}, {Cros}, {Denby},
  {Dhez}, {Di Coco}, {Dowson}, {Ferrando}, {Ghizzardi}, {Gianotti}, {Goodall},
  {Gretton}, {Griffiths}, {Hainaut}, {Hochedez}, {Holland}, {Jourdain},
  {Kendziorra}, {Lagostina}, {Laine}, {La Palombara}, {Lortholary}, {Lumb},
  {Marty}, {Molendi}, {Pigot}, {Poindron}, {Pounds}, {Reeves}, {Reppin},
  {Rothenflug}, {Salvetat}, {Sauvageot}, {Schmitt}, {Sembay}, {Short},
  {Spragg}, {Stephen}, {Str{\"u}der}, {Tiengo}, {Trifoglio}, {Tr{\"u}mper},
  {Vercellone}, {Vigroux}, {Villa}, {Ward}, {Whitehead}, \& {Zonca}}]{tur01}
{Turner}, M.~J.~L., {Abbey}, A., {Arnaud}, M., {et~al.} 2001, \aap, 365, L27

\bibitem[{{Villata} {et~al.}(2007){Villata}, {Raiteri}, {Aller}, {Bach},
  {Ibrahimov}, {Kovalev}, {Kurtanidze}, {Larionov}, {Lee}, {Leto},
  {L{\"a}hteenm{\"a}ki}, {Nilsson}, {Pursimo}, {Ros}, {Sumitomo}, {Volvach},
  {Aller}, {Arai}, {Buemi}, {Coloma}, {Doroshenko}, {Efimov}, {Fuhrmann},
  {Hagen-Thorn}, {Kamada}, {Katsuura}, {Konstantinova}, {Kopatskaya}, {Kotaka},
  {Kovalev}, {Kurosaki}, {Lanteri}, {Larionova}, {Mingaliev}, {Mizoguchi},
  {Nakamura}, {Nikolashvili}, {Nishiyama}, {Sadakane}, {Sergeev}, {Sigua},
  {Sillanp{\"a}{\"a}}, {Smart}, {Takalo}, {Tanaka}, {Tornikoski}, {Trigilio},
  \& {Umana}}]{vil07}
{Villata}, M., {Raiteri}, C.~M., {Aller}, M.~F., {et~al.} 2007, \aap, 464, L5

\bibitem[{{Villata} {et~al.}(2006){Villata}, {Raiteri}, {Balonek}, {Aller},
  {Jorstad}, {Kurtanidze}, {Nicastro}, {Nilsson}, {Aller}, {Arai}, {Arkharov},
  {Bach}, {Ben{\'{\i}}tez}, {Berdyugin}, {Buemi}, {B{\"o}ttcher}, {Carosati},
  {Casas}, {Caulet}, {Chen}, {Chiang}, {Chou}, {Ciprini}, {Coloma}, {di Rico},
  {D{\'{\i}}az}, {Efimova}, {Forsyth}, {Frasca}, {Fuhrmann}, {Gadway}, {Gupta},
  {Hagen-Thorn}, {Harvey}, {Heidt}, {Hernandez-Toledo}, {Hroch}, {Hu}, {Hudec},
  {Ibrahimov}, {Imada}, {Kamata}, {Kato}, {Katsuura}, {Konstantinova},
  {Kopatskaya}, {Kotaka}, {Kovalev}, {Kovalev}, {Krichbaum}, {Kubota},
  {Kurosaki}, {Lanteri}, {Larionov}, {Larionova}, {Laurikainen}, {Lee}, {Leto},
  {L{\"a}hteenm{\"a}ki}, {L{\'o}pez-Cruz}, {Marilli}, {Marscher}, {McHardy},
  {Mondal}, {Mullan}, {Napoleone}, {Nikolashvili}, {Ohlert}, {Postnikov},
  {Pursimo}, {Ragni}, {Ros}, {Sadakane}, {Sadun}, {Savolainen}, {Sergeeva},
  {Sigua}, {Sillanp{\"a}{\"a}}, {Sixtova}, {Sumitomo}, {Takalo},
  {Ter{\"a}sranta}, {Tornikoski}, {Trigilio}, {Umana}, {Volvach}, {Voss}, \&
  {Wortel}}]{vil06}
{Villata}, M., {Raiteri}, C.~M., {Balonek}, T.~J., {et~al.} 2006, \aap, 453,
  817

\bibitem[{{Villata} {et~al.}(2004){Villata}, {Raiteri}, {Kurtanidze},
  {Nikolashvili}, {Ibrahimov}, {Papadakis}, {Tosti}, {Hroch}, {Takalo},
  {Sillanp{\"a}{\"a}}, {Hagen-Thorn}, {Larionov}, {Schwartz}, {Basler},
  {Brown}, {Balonek}, {Ben{\'{\i}}tez}, {Ram{\'{\i}}rez}, {Sadun}, {Boltwood},
  {Carini}, {Barnaby}, {Coloma}, {Ros}, {Dai}, {Xie}, {Mattox}, {Rodriguez},
  {Asfandiyarov}, {Atkerson}, {Beem}, {Bloom}, {Chanturiya}, {Ciprini},
  {Crapanzano}, {de Diego}, {Efimova}, {Gardiol}, {Guerra}, {Kahharov},
  {Kapanadze}, {Karttunen}, {Kato}, {Kimeridze}, {Kudryavtseva}, {Lainela},
  {Lanteri}, {Larionova}, {Maesano}, {Marchili}, {Massone}, {Monroe},
  {Montagni}, {Nesci}, {Nilsson}, {Noble}, {Nucciarelli}, {Ostorero},
  {Papamastorakis}, {Pasanen}, {Peters}, {Pursimo}, {Reig}, {Ryle}, {Sclavi},
  {Sigua}, {Uemura}, \& {Wills}}]{vil04a}
{Villata}, M., {Raiteri}, C.~M., {Kurtanidze}, O.~M., {et~al.} 2004, \aap, 421,
  103

\bibitem[{{Villata} {et~al.}(2002){Villata}, {Raiteri}, {Kurtanidze},
  {Nikolashvili}, {Ibrahimov}, {Papadakis}, {Tsinganos}, {Sadakane}, {Okada},
  {Takalo}, {Sillanp{\"a}{\"a}}, {Tosti}, {Ciprini}, {Frasca}, {Marilli},
  {Robb}, {Noble}, {Jorstad}, {Hagen-Thorn}, {Larionov}, {Nesci}, {Maesano},
  {Schwartz}, {Basler}, {Gorham}, {Iwamatsu}, {Kato}, {Pullen},
  {Ben{\'{\i}}tez}, {de Diego}, {Moilanen}, {Oksanen}, {Rodriguez}, {Sadun},
  {Kelly}, {Carini}, {Miller}, {Catalano}, {Dultzin-Hacyan}, {Fan}, {Ishioka},
  {Karttunen}, {Kein{\"a}nen}, {Kudryavtseva}, {Lainela}, {Lanteri},
  {Larionova}, {Matsumoto}, {Mattox}, {Montagni}, {Nucciarelli}, {Ostorero},
  {Papamastorakis}, {Pasanen}, {Sobrito}, \& {Uemura}}]{vil02}
{Villata}, M., {Raiteri}, C.~M., {Kurtanidze}, O.~M., {et~al.} 2002, \aap, 390,
  407

\bibitem[{{von Montigny} {et~al.}(1997){von Montigny}, {Aller}, {Aller},
  {Bruhweiler}, {Collmar}, {Courvoisier}, {Edwards}, {Fichtel}, {Fruscione},
  {Ghisellini}, {Hartman}, {Johnson}, {Kafatos}, {Kii}, {Kniffen}, {Lichti},
  {Makino}, {Mannheim}, {Marscher}, {McBreen}, {McHardy}, {Pesce}, {Pohl},
  {Ramos}, {Reich}, {Robson}, {Sasaki}, {Teraesranta}, {Tornikoski}, {Urry},
  {Valtaoja}, {Wagner}, \& {Weekes}}]{von97}
{von Montigny}, C., {Aller}, H., {Aller}, M., {et~al.} 1997, \apj, 483, 161

\bibitem[{{Wills}(1991)}]{wil91}
{Wills}, B.~J. 1991, in Variability of Active Galactic Nuclei, ed. H.~R.
  {Miller} \& P.~J. {Wiita}, 87--101

\bibitem[{{Wills} {et~al.}(1985){Wills}, {Netzer}, \& {Wills}}]{wil85}
{Wills}, B.~J., {Netzer}, H., \& {Wills}, D. 1985, \apj, 288, 94

\bibitem[{{Wilms} {et~al.}(2000){Wilms}, {Allen}, \& {McCray}}]{wil00}
{Wilms}, J., {Allen}, A., \& {McCray}, R. 2000, \apj, 542, 914

\bibitem[{{Wolfe} {et~al.}(1982){Wolfe}, {Briggs}, \& {Davis}}]{wol82}
{Wolfe}, A.~M., {Briggs}, F.~H., \& {Davis}, M.~M. 1982, \apj, 259, 495

\bibitem[{{Wu} {et~al.}(2007){Wu}, {Zhou}, {Ma}, {Wu}, {Jiang}, \&
  {Chen}}]{wu07}
{Wu}, J., {Zhou}, X., {Ma}, J., {et~al.} 2007, \aj, 133, 1599

\end{thebibliography}
\end{document}